\documentclass[]{scrartcl}

\usepackage{authblk}

\usepackage{microtype}
\usepackage[bibstyle=numeric,citestyle=numeric, uniquename=false,sortcites=true,maxbibnames=20, maxcitenames=20,doi=false, url=true, eprint=false, firstinits=true, backend=bibtex, natbib=true, isbn=false, date=year]{biblatex} 
\bibliography{litATpaper}
\usepackage{amsmath}
\usepackage{amssymb}
\usepackage{mathtools}
\usepackage{textcomp,eurosym}
\usepackage{booktabs}

\usepackage{url}
\usepackage{graphicx}
\usepackage{floatflt}
\usepackage{dblfloatfix, wrapfig}
\usepackage{color}
\usepackage[algo2e]{algorithm2e}
\usepackage{footnote}
\makesavenoteenv{tabular}

\usepackage{units}

\usepackage[T1]{fontenc}
\usepackage{lmodern}

\newcommand{\nodState}{x}
\newcommand{\nodInput}{u}
\newcommand{\nodDist}{d}

\definecolor{darkblue}{rgb}{0,0,0.9}
\definecolor{orange}{rgb}{1,0.5,0}
\definecolor{htmlgreen}{rgb}{0.0, 0.5, 0.0}

\usepackage{xspace}
\newcommand{\aladin}{\textsc{aladin}\xspace}
\newcommand{\admm}{\textsc{admm}\xspace}
\newcommand{\nlp}{\textsc{nlp}\xspace}

\newcommand{\opf}{\textsc{opf}\xspace}
\newcommand{\sopf}{s\textsc{opf}\xspace}
\newcommand{\ac}{\textsc{ac}\xspace}
\newcommand{\dc}{\textsc{dc}\xspace}

\newcommand{\sdp}{\textsc{sdp}\xspace}
\newcommand{\ieee}{\textsc{ieee}\xspace}
\newcommand{\sqp}{\textsc{sqp}\xspace}
\newcommand{\tso}{\textsc{tso}\xspace}
\newcommand{\dso}{\textsc{dso}\xspace}
\newcommand{\eex}{\textsc{eex}\xspace}
\newcommand{\res}{\textsc{res}\xspace}
\newcommand{\scada}{\textsc{scada}\xspace}
\newcommand{\facts}{\textsc{facts}\xspace}
\newcommand{\nmpc}{\textsc{nmpc}\xspace}

\newcommand{\qp}{\textsc{qp}\xspace}

\newcommand{\mbb}[1]{\mathbb #1}

\newcommand{\mcl}[1]{\mathcal #1}

\newcommand{\imag}{\mathrm{j}}

\newcommand{\nodStateDC}{x^{\text{\dc}}}
\newcommand{\nodInputDC}{u^{\text{\dc}}}
\newcommand{\nodDistDC}{d^{\text{\dc}}}

\newcommand{\ProbMeasure}{\mathbb{P}}
\newcommand{\Outcomes}{\Omega}
\newcommand{\Ltwospace}[1]{L^2(\Outcomes, \ProbMeasure; \mathbb{R}^{#1})}
\newcommand{\ProbSpace}{(\Outcomes, \mathcal{A}, \ProbMeasure)}
\newcommand{\rv}[1]{\mathsf{#1}}
\newcommand{\ev}[1]{\mathrm{E}_{\ProbMeasure}\left[ #1 \right]}

\def\keywords{\vspace{0.1em}
	{\textit{Keywords}:\,\relax%
}}

\begin{document}

\author{Timm Faulwasser}
\author{Alexander Engelmann}
\author{Tillmann M{\"u}hlpfordt} 
\author{Veit Hagenmeyer}

\affil{Institute for Automation and Applied Informatics (IAI) \\ 
	   Karlsruhe Institute of Technology (KIT)\\ Hermann-von-Helmholtz-Platz 1 \\
	   76344 Eggenstein-Leopoldshafen, Germany \\
	E-Mail: timm.faulwasser@ieee.org,\\$\{$alexander.engelmann, tillmann.muehlpfordt, veit.hagenmeyer$\}$@kit.edu
}

 \title{Optimal Power Flow: \\
	An Introduction to Predictive, Distributed and Stochastic Control Challenges}
\date{}

\maketitle

\begin{abstract}
The \emph{Energiewende} is a paradigm change that can be witnessed at latest since the political decision to step out of nuclear energy. Moreover, despite common roots in Electrical Engineering, the control community and the power systems community face a lack of common vocabulary.
In this context, this paper aims at providing a systems-and-control specific introduction to optimal power flow problems which are pivotal in the operation of energy systems.
Based on a concise problem statement, we introduce a common description of  optimal power flow variants including multi-stage-problems and predictive control, stochastic uncertainties, and issues of distributed optimization. Moreover, we sketch open  questions that might be of interest for the systems and control community. 
\end{abstract}
 \keywords{Optimal power flow, stochastic uncertainties, distributed optimization, model predictive control}

\section{Introduction}  \label{sec:intro}
The \emph{Energiewende} is not an event of the distant future; rather it is a paradigm change whose matter-of-factness we witness at latest since the political post-Fukushima decision of the German government to step out of nuclear energy on the of July 1, 2011.  The same day, the German government took the decision to raise the share of so-called Renewable Energy Sources (\res) in the electricity sector up to 80\% by 2050, and decided to invest massively in the extension of the transmission grid \cite{Bundestag_Atomausstieg:11}.  Important dimensions and consequences of this transition---which, actually and in the  view of greenhouse gas emission reduction  to counteract global warming, affects not only Germany but a large number of countries world-wide---include the change from a rather small number of large-scale power plants acting on the high-voltage transmission level towards a large number of small-scale acting pre-dominantly on the medium to low-voltage distribution level \cite{GFMEAE2014}. 

Thus, from a control and automation point of view, the wide range of research needs and challenges induced by the Energiewende entails:
\begin{itemize}
\item Development of new power systems hardware, grid-aware market design,  flexible IT solutions, and \scada systems overcoming the monolithic structure of current process automation systems \cite{Farhangi2010,Buchholz2013}.
\item Investigation of new decentralized and distributed control methods ensuring stable operation of distribution grids with a large share of volatile/uncertain renewables (which implies a large number of controllable devices such as storages and devices of sector coupling) and allowing for adaptive changes of grid topology (islanding of subsystems/microgrids) \cite{Olivares2014,Kroposki2017,Benz2015,kit:braun18a}.
\item Tailoring numerical algorithms that contribute to safe, secure and economically efficient operation of high-voltage transmission grids, including the structured consideration of uncertainties (volatile renewables etc.) \cite{Bienstock2014,Low2014,Peng2018}.
\end{itemize}
We remark that neither does the above list claim completeness nor does the order imply any prioritization.

In this context, it is worth to be noted that, despite common roots in Electrical Engineering, the control community and the power systems community face a lack of common vocabulary. The frequently cited---if not seminal---task-force paper on different stability notions in power systems and in systems and control is a prime evidence of the lack of common vocabulary \cite{Kundur04a}. Furthermore,  the lack of widespread knowledge about the importance of optimal power flow problems in the control community can be regarded as another evidence.

\begin{figure}[t]
	\centering
	\includegraphics[width=0.35\textwidth]{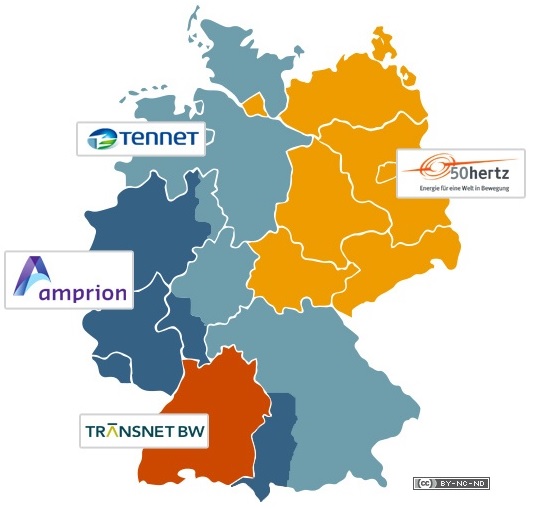} \includegraphics[width=0.25\textwidth]{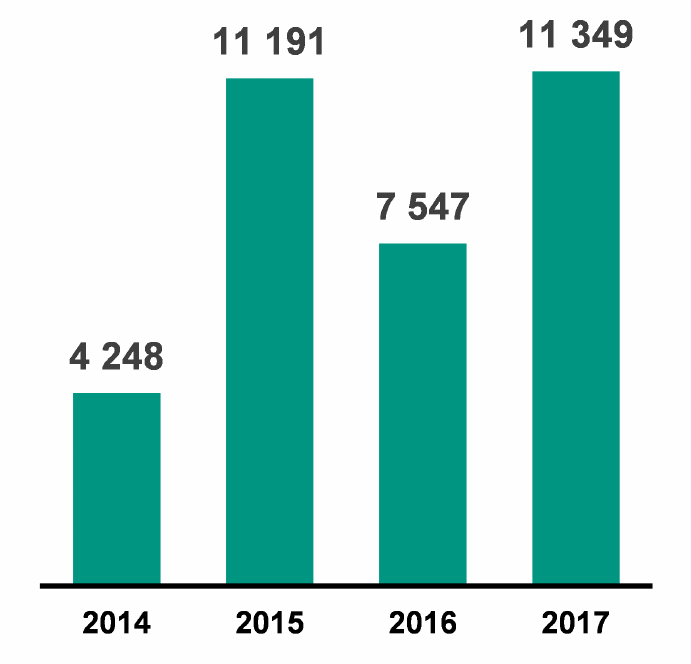}
	\caption{(Left) Control areas of German \tso{}s. Source: \emph{Bundeszentrale f\"ur politische Bildung, 2013, www.bpb.de}. (Right) Annual re-dispatch (in GWh) of German \tso{}s. Data source: \emph{netztransparenz.de}}
	\label{fig:TSO}
\end{figure}

On a macroscopic level, the present paper aims at contributing towards consistent notions in power and control systems engineering. On a detailed level, we will not touch upon the frequently discussed problems of smart-grid control and modeling \cite{Schiffer16a, Riverso18a}. Rather,  we focus on the so-called Optimal Power Flow (\opf) Problem, which arises in different contexts of operation of electricity grids. One may say that \opf is \emph{the} most important steady-state optimization problem arising in power systems. For example, it plays a pivotal role in computing setpoints for grid stabilizing generators whenever the market solution at the European Energy Exchange (\eex) in Leipzig \cite{EUROPEANCOMMISSION2010} is incompatible with the physics of the grid.    
In other words, whenever the market solution might jeopardize grid stability, the  Transmission System Operators (\tso{}s) ramp-up back-up power plants whose setpoints are determined by solving \opf problems. Figure~\ref{fig:TSO} depicts the control regions of the four German \tso{}s (left) and the annual amount of energy re-dispatched by the four German \tso{}s (right) in 2014-2017. Evidently, with about \unit[11]{TWh} the annual re-dispatch has reached a level inducing substantial economic costs. For example, in 2015 the re-dispatch costs added up to \unit[402.5]{Million \euro} \cite{BGGA2016}.\footnote{This is one of the reasons for the steep increase in electricity prices in Germany from \unit[21.07]{cent/kWh} in 2007 to  \unit[30.48]{cent/kWh} in 2017  \cite{EUROSTAT2017}.} 

Within the usual time interval of \unit[15]{min} a power plant can change its setpoint only incrementally. Thus, in real-world applications (multi-stage quasi-stationary) \opf problems are subject to hidden constraints that can be expressed as discrete-time dynamics. Put differently,  \opf leads naturally to large-scale non-convex discrete-time optimal control problems. Moreover, as the Energiewende induces uncertainties (volatile renewables, uncertain future market solutions, ...) there is a tremendous need for scalable methods tackling \opf problems.  Hence, in the present paper we aim at providing a unified framework to  multi-stage predictive, distributed and stochastic \opf. Moreover, the paper is meant as an introduction  tailored to readers with background in systems and control who are not yet familiar with \opf. This scope implies that we do neither claim completeness in terms of literature overview, nor do we discuss the most general variants of \opf problems. 

We begin with a concise statement of the \opf problem in Section \ref{sec:Problem} commenting on usual relaxations. Section \ref{sec:Challenges} entails the main contribution of this paper; i.e. a control-specific formulation of research challenges all of which entail \opf problems and variants thereof at their core. This includes blueprint formulations for multi-stage quasi-stationary predictive \opf, distributed (multi-agent) formulations,  \opf with stochastic uncertainties, and comments on further \opf variants. 
Moreover, in Section \ref{sec:openProbs} we sketch control-specific open problems. Finally, the paper closes with conclusions in Section \ref{sec:Conclusions}.
\section{Optimal Power Flow -- Problem Statement} \label{sec:Problem}
There exists a plethora of references on \opf, see \cite{Zhu2015, Das2017,Frank2016,Carpentier1962, Capitanescu2016}. Subsequently, we present a concise formulation of \opf problems that enables the statement of control-specific research challenges in Section \ref{sec:Challenges}.

\subsection{The Power Flow Equations} \label{sec:PowerFlow}
We consider balanced electrical \ac grids as lumped-parameter systems at steady state, which can be modeled by the triple $(\mathcal{N},\mathcal{G},Y)$, where $\mathcal{N}=\{1, \hdots, N\}$ is the set of buses (nodes), $\mathcal{G} \subseteq \mathcal{N}$ is the non-empty set of generators, and $Y = G + \imag B \in \mathbb{C}^{N\times N}$ is the bus admittance matrix \cite{Grainger1994}. 
Moreover, we assume symmetric three-phase \ac conditions.  
Every bus $l \in \mathcal{N}$ is described by its voltage phasor $v_l \mathrm{e}^{j \theta_l} \in \mathbb{C}$ and net apparent power $s_l = p_l + j q_l \in \mathbb{C}$, or equivalently by its voltage magnitude $v_l$, voltage phase $\theta_l$, net active power $p_l$, and net reactive power $q_l$.

\subsubsection{AC Power Flow}
\begin{figure}[b]
	\centering
	\includegraphics[width=0.3\textwidth]{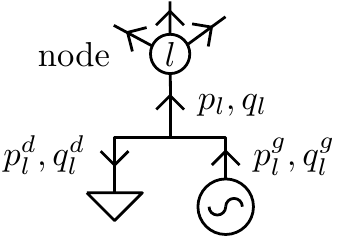}
	\caption{Power generation, demand and net power at node $l$.}
	\label{fig:NetPower}
\end{figure}

The power flow equations describe the steady-state behavior of an \ac electrical network in terms of the voltage phasors and net apparent powers
\begin{subequations}
	\label{eq:powerflow}
	\begin{align}
	p_l = &\;v_l\sum_{m \in \mathcal{N}} v_m(G_{lm}\cos(\theta_{lm})+B_{lm}\sin(\theta_{lm})),\\
	q_l = &\;v_l\sum_{m \in \mathcal{N}} v_m(G_{lm}\sin(\theta_{lm})-B_{lm}\cos(\theta_{lm})),
	\end{align}
\end{subequations}
where $\theta_{lm} := \theta_l -\theta_m$.
Observe that in the  power flow equations~\eqref{eq:powerflow} the phase angles $\theta_l$ occur as pair-wise differences, therefore one bus $l_0\in \mathcal{N}$ is specified as reference (slack) bus 
$	\theta_{l_0} = 0$ for $	l_0 \in \mcl{N}$; w.l.o.g. we consider $l_0 = 1$ in the remainder.
For the sake of simplicity, we assume that there is only one generator per bus (i.e. $\mcl{G} \subseteq \mcl{N}$). We describe the net apparent power of bus $l \in \mathcal{N}$ by
\begin{equation*}
s_l = p_l + \imag q_l =
\begin{cases}
(p_l^{\text{g}} - p_{l}^{\text{d}}) + \imag ( q_{l}^{\text{g}} - q_{l}^{\text{d}} ), & l \in \mathcal{G}, \\
- p_{l}^{\text{d}} - \imag q_{l}^{\text{d}}, & \text{otherwise},
\end{cases}
\end{equation*}
where $p_{l}^{\text{g}}$, $q_{l}^{\text{g}}$ are controllable power injections for all generator nodes $l \in \mathcal{G}$, and $p_{l}^{\text{d}}$, $q_{l}^{\text{d}}$ are uncontrollable power sinks/sources for all $l \in \mathcal{N}$, cf. Figure~\ref{fig:NetPower}.

Hence, we define the control input $\nodInput \in \mathbb{R}^{n_{\nodInput}}$, the disturbance $\nodDist \in \mathbb{R}^{n_{\nodDist}}$, and the state $\nodState \in \mathbb{R}^{n_{\nodState}}$ as follows
\begin{subequations}
\begin{align} 
\nodInput &=	\begin{pmatrix} p_{l}^{\text{g}} & q_{l}^{\text{g}}\end{pmatrix}_{l\in\mcl{G}}^\top \in \mbb{R}^{n_{\nodInput}},  && n_{\nodInput} = 2 | \mathcal{G} |, \label{eq:NodInput} \\
\nodDist &=	\begin{pmatrix} p_{l}^{\text{d}}& q_{l}^{\text{d}} \end{pmatrix}^\top_{l\in\mcl{N}}  \in \mbb{R}^{n_{\nodDist}}, && n_{\nodDist} = 2 | \mcl{N} |, \label{eq:NodDist}\\
\nodState &= \begin{pmatrix}v_l & \theta_l \end{pmatrix}_{l\in\mcl{N}}^\top \in \mbb{R}^{n_{\nodState}},  && n_{\nodState} = 2 | \mcl{N} |. \label{eq:NodState} 
\end{align}
\end{subequations}
We remark that depending on the specific problem at hand, one may also consider the voltage of a bus as an additional input variable. 
Our choice of state and input variables allows writing the power flow equations~\eqref{eq:powerflow} in terms of a system of nonlinear algebraic equations
\begin{equation}
\label{eq:PFE_compact}
F: \mbb{R}^{n_{\nodState}} \times \mbb{R}^{n_{\nodInput}} \times \mbb{R}^{n_{\nodDist}} \rightarrow \mbb{R}^{2 N} \qquad  F(\nodState,\nodInput; \nodDist) = 0,
\end{equation}
where the semicolon notation emphasizes the dependency on the exogenous disturbance $\nodDist$.
For the sake of concise notation, we introduce the so-called power-flow manifold
\begin{equation} \label{eq:PFEconstrSet}
\mcl{F}(\nodDist) := \left\{(\nodState ~ \nodInput)^\top \in \mathbb{R}^{n_{\nodState} + n_{\nodInput}} \,|\, F( \nodState, \nodInput; \nodDist) = 0  \right\} 
\end{equation}
describing all solutions to the power-flow equations \eqref{eq:powerflow} for a given disturbance $\nodDist$.
Since \eqref{eq:powerflow} states $2N$ equality constraints and we consider $4N$ variables, the differentiable manifold $\mcl{F}(\nodDist)$ is of dimension $2N$. As we will see later, it determines the number of degrees of freedom remaining for optimization.

Note that in this section we have formulated the power-flow equations in polar coordinates.
However, resorting to Cartesian coordinates---i.e. swapping voltage magnitude $v_l$ and phase $\theta_l$ with the imaginary and real part of the voltage---they can equivalently be written as a set of polynomial equations \cite{Frank2016}.

\subsubsection{DC Power Flow}
To the end of obtaining a linear approximation of the power flow equations \eqref{eq:powerflow}, one typically assumes the following:
lossless lines ($r_{lm} \approx 0$ for the Ohmic resistance of the line connecting buses $l$ and $m$), small phase differences ($\theta_{lm} = \theta_l - \theta_m \approx 0$), constant voltage magnitudes ($v_l \approx 1$).
Under these assumptions---which typically hold for high-voltage transmission systems---the \ac active power flow equations \eqref{eq:powerflow} simplify to
\begin{align}
\label{eq:DCpowerflow}
p_l  = &\; - \sum_{m \in \mathcal{N} \setminus \{ l\} } b_{lm} (\theta_l - \theta_m) \quad \Longleftrightarrow \quad p  = - B \theta,
\end{align}
where $B$ is the imaginary part of the bus admittance matrix $Y$.
The above equations~\eqref{eq:DCpowerflow} are the so-called \dc power flow equations \cite{Grainger1994}.
Note that in the absence of shunts elements the reactive power flows vanish due to the assumptions of small angle differences and constant voltages.
Compared to \ac power flow, the control input $\nodInputDC \in \mathbb{R}^{n_{\nodInputDC}}$, the disturbance $\nodDistDC \in \mathbb{R}^{n_{\nodDistDC}}$, and the state $\nodStateDC \in \mathbb{R}^{n_{\nodStateDC}}$ for \dc power flow become
\begin{subequations}
	\begin{align*} 
	\nodInputDC &=	\begin{pmatrix} p_{l}^{\text{g}} \end{pmatrix}_{l\in\mcl{G}}^\top \in \mbb{R}^{n_{\nodInputDC}},  && n_{\nodInputDC} = | \mathcal{G} |, \\
	\nodDistDC &=	\begin{pmatrix} p_{l}^{\text{d}} \end{pmatrix}^\top_{l\in\mcl{N}}  \in \mbb{R}^{n_{\nodDistDC}}, && n_{\nodDistDC} = | \mcl{N} |, \\
	\nodStateDC &= \begin{pmatrix} \theta_l \end{pmatrix}_{l\in\mcl{N}}^\top \in \mbb{R}^{n_{\nodStateDC}},  && n_{\nodStateDC} = | \mcl{N} |.
	\end{align*}
\end{subequations}
From this it follows that the net power $p$ is $p = \nodInputDC + \nodDistDC$.
In terms of the compact notation~\eqref{eq:PFE_compact} and \eqref{eq:PFEconstrSet} we obtain
\begin{equation*}
F^{\text{\dc}}(\nodStateDC, \nodInputDC; \nodDistDC) = \nodInputDC - \nodDistDC + B \nodStateDC = 0,
\end{equation*}
and
\begin{equation} \label{eq:PFEconstrSet_DC}
\mcl{F}^{\text{\dc}}(\nodDistDC) := \left\{(\nodStateDC ~ \nodInputDC)^\top \in \mathbb{R}^{n_{\nodStateDC} + n_{\nodInputDC}} \,|\, F^{\text{\dc}}(\nodStateDC, \nodInputDC; \nodDistDC) = 0  \right\}.
\end{equation}

\subsection{Optimal Power Flow} \label{sec:AC-OPF}
Besides the power flow equations \eqref{eq:powerflow} engineering requirements are usually considered in terms of box constraints as follows
\begin{subequations}
	\label{eq:Constraints_StateInput}
	\begin{align}
	\nodInput \in \mcl{U} &= \left\{ \begin{pmatrix} p_{l}^{\text{g}} & q_{l}^{\text{g}}\end{pmatrix}_{l\in\mcl{G}}^\top \in \mbb{R}^{n_u}: \quad   p_{l}^{\text{g}} \in [\underline{p}_{l}^{\text{g}}\,, \overline{p}_{l}^{\text{g}}], \quad q_{l}^{\text{g}} \in [ \underline{q}_{l}^{\text{g}},\, \overline{q}_{l}^{\text{g}}] \right\} \\
	\nodState \in \mcl{X} &= \left\{ \begin{pmatrix}v_l & \theta_l \end{pmatrix}_{l\in\mcl{N}}^\top \in \mbb{R}^{n_x}: \quad  v_l \in [ \underline{v}_l,\, \overline{v}_l], \: \forall l \in \mcl{N}, \quad \theta_{l_0} = 0 \right\}.
	\end{align}
\end{subequations}
Actually, the constraints~\eqref{eq:Constraints_StateInput} are a simplification of the true technical requirements.
For example, generator curves may impose constraints that couple $p_{l}^{\text{g}}$, $q_{l}^{\text{g}}$, and $v_l$ at generator bus $l \in \mcl{G}$;
binary constraints may be imposed when shunts and/or generators can be turned on and off.

Additional constraints comprise limits on the line flows, which are often modeled as constraints on the magnitude of apparent power
\begin{subequations} \label{eq:congestions}
\begin{equation}
| s_{lm} | = \sqrt{p_{lm}^2 + q_{lm}^2} \leq |\bar{s}_{lm}|, \quad \forall (l,m) \in \mcl{L},
\end{equation}
where $\mcl{L} \subseteq \mcl{N} \times \mcl{N}$ is the set of lines, with $M = |\mathcal{L}|$ being the number of lines.
The active power $p_{lm}$ across the line connecting buses $l$ to $m$, $p_{lm}$, and the corresponding reactive power $q_{lm}$ are given by
\begin{align}
p_{lm}&=\hphantom{-} v_l^2 g_{lm} - v_{l} v_{m} (g_{lm} \cos(\theta_{lm}) + b_{lm}\sin(\theta_{lm})), \\
q_{lm}&=-v_l^2 b_{lm} + v_{l} v_{m} (b_{lm} \cos(\theta_{lm}) - g_{lm}\sin(\theta_{lm})),
\end{align}
 \end{subequations}
where $b_{lm}$ and $g_{lm}$ are line parameters. 
The line flows \eqref{eq:congestions} depend only on the state $\nodState$, and can be written compactly as
\begin{equation}
\label{eq:LineFlowConstraints}
c: \mbb{R}^{n_{\nodState}} \rightarrow \mbb{R}^{|\mcl{L}|} \qquad  \qquad s_{\text{line}} = c(\nodState).
\end{equation}
This allows for the compact notation
 \begin{equation} \label{eq:lineConstrSet}
 \mcl{C} := \left\{\nodState \in \mbb{R}^{n_{\nodState}} \,|\, c(x) \leq \begin{pmatrix}
|\bar{s}_{lm}| 
 \end{pmatrix}_{(l,m) \in \mcl{L}}^\top \right\} \subset \mbb{R}^{n_{\nodState}},
 \end{equation}
where the inequality is evaluated component-wise.

\subsubsection{AC OPF}
Typical objectives considered in \opf span from the minimization of active power generation costs via the minimization of transmission losses to suppressing overly large injections of reactive power, see e.g. \cite{Frank2016}.
Frequently, the cost function $J: \mathbb{R}^{n_{\nodInput}} \rightarrow \mathbb{R}$ is assumed to be convex (often quadratic) in the argument $\nodInput$.
Summarizing all of the above, the single-stage \ac \opf problem is given by
\begin{subequations}
	\label{eq:AC_OPF}
	\begin{align} 
	\min_{(\nodState, \nodInput)\in \mathbb{R}^{n_{\nodState}+n_{\nodInput}}} \quad & J(\nodInput)  \\
	\text{subject to}&  \notag\\
	(\nodState, ~ \nodInput)^\top &\in \mcl{F}(\nodDist), \\
		\nodInput &\in \mcl{U}, \\
	\nodState &\in \mcl{X},  \label{eq:ACOPF_convex}\\
	\nodState & \in \mcl{C}.
	\end{align}
\end{subequations}
The main challenges in solving the \ac \opf as given above are: the non-convexity of the sets $ \mcl{F}(\nodDist)$ and $ \mcl{C}$, the fact that the objective is not strictly convex in $\nodState$ \emph{and} $\nodInput$, and the fact that realistic grid models can easily comprise several thousand nodes \cite{Semerow2015,Villella2012}. 

Note that there exist powerful software packages to solving \eqref{eq:AC_OPF} such as \textsc{Matpower} \cite{Zimmerman11} (which is an open-source \textsc{Matlab} toolbox specific for \opf), \textsc{jump} \cite{Dunning17a} (which is an open-source \textsc{Julia} \nlp package) or \textsc{CasADi} \cite{Andersson12a} (which is an open-source  \nlp package available for \textsc{Matlab} and \textsc{Python}).
In general, one can distinguish the following main approaches to solving \eqref{eq:AC_OPF}:
\begin{itemize}
\item[(i)] tackling it as a generic \nlp \cite{Carpentier1962,Momoh1999,Zhu2015,Frank2016}; 
\item[(ii)] alternating  solution of the power-flow equations \eqref{eq:powerflow} and minimizing a quadratic objective \cite{Hauswirth2017,Salgado1990}; 
\item[(iii)] polynomial reformulation and semi-definite relaxation of the power-flow equations \eqref{eq:powerflow} \cite{Bai2008,Low2014,Low2014a,Mehta2016}; and
\item[(iv)] affine approximation of the sets $\mcl{F}(\nodDist)$ and $\mcl{C}$ which is discussed next.
\end{itemize}

\subsubsection{DC OPF} \label{sec:DC-OPF}
For \dc power flow conditions, the box constraints for the input $\nodInputDC$ and the state $\nodStateDC$ are obtained from \eqref{eq:Constraints_StateInput} by removing the entries referring to the reactive power and the voltage magnitude respectively
\begin{subequations}
	\label{eq:Constraints_StateInput_DC}
	\begin{align*}
	\nodInputDC \in \mcl{U}^{\text{\dc}} &:= \left\{ \begin{pmatrix} p_{l}^{\text{g}} \end{pmatrix}_{l\in\mcl{G}}^\top \in \mbb{R}^{n_u}: \quad   p_{l}^{\text{g}} \in [\underline{p}_{l}^{\text{g}}, \overline{p}_{l}^{\text{g}}] \quad \forall l \in \mcl{G} \right\} \\
	\nodStateDC \in \mcl{X}^{\text{\dc}} &:= \left\{ \begin{pmatrix} \theta_l \end{pmatrix}_{l\in\mcl{N}}^\top \in \mbb{R}^{n_x}: \quad  \theta_{l_0} = 0 \right\}.
	\end{align*}
\end{subequations}
Under \dc conditions the function $c^{\text{\dc}}: \mbb{R}^{n_{\nodStateDC}} \rightarrow \mathbb{R}^{| \mcl{L} |}$ in  \eqref{eq:LineFlowConstraints}---that maps the state to the line flows---is obtained from Kirchhoff's laws as $c^{\text{\dc}}(\nodStateDC) = - B_{\text{br}} A \nodStateDC$, where $B_{\text{br}} = \operatorname{diag} ( (b_{lm})_{(l,m) \in \mathcal{L}}^{\top}  ) \in \mathbb{R}^{M \times M} $ is the branch susceptance matrix, and $A \in \mathbb{R}^{M \times N}$ is the graph incidence matrix.
This leads to
\begin{equation*}
\mathcal{C}^{\text{\dc}} := \left\{ \nodStateDC \in \mathbb{R}^{n_{\nodStateDC}} | -B_{\text{br}} A \nodStateDC \leq \begin{pmatrix}
\bar{p}_{lm}
\end{pmatrix}_{(l,m) \in \mcl{L}}^\top \right\} \subset \mbb{R}^{n_{\nodStateDC}}.
\end{equation*}
Summing up, the single-stage \dc \opf problem can be posed as follows
\begin{subequations}
	\label{eq:DC_OPF}
	\begin{align} 
	\min_{(\nodStateDC,\, \nodInputDC)\in \mathbb{R}^{n_{\nodStateDC}+n_{\nodInputDC} }} \quad & J(\nodInputDC)  \\
	\text{subject to}&  \notag\\
	(\nodStateDC, ~ \nodInputDC)^\top &\in \mcl{F}^{\text{\dc}}(\nodDistDC), \\
	\nodInputDC &\in \mcl{U}^{\text{\dc}}, \\
	\nodStateDC &\in \mcl{X}^{\text{\dc}}, \\
	\nodStateDC &\in \mcl{C}^{\text{\dc}}.
	\end{align}
\end{subequations}
Observe that, for usual choices of $J$, \eqref{eq:DC_OPF} is a quadratic program positive definite in $\nodInputDC$.  Hence \dc \opf is structurally considerably simpler than \ac \opf.
Finally, note that it is straightforward to eliminate the state $\nodStateDC$ from \eqref{eq:DC_OPF}. This yields smaller optimization problems that are strictly convex in the decision variable $\nodInputDC$.

\section{Advanced OPF Variants } \label{sec:Challenges}

After the introduction of the \opf problem in its simple single-stage \ac and \dc variants, it deserves to be noted that in the power systems community these problems are for the most part well understood, see \cite{Grainger1994,Momoh1999,Zhu2015,Frank2016}.\footnote{One remaining open issue is, for example, the one of uniqueness of solutions \cite{Nick2018}.} At the same time Problem \eqref{eq:AC_OPF} and Problem \eqref{eq:DC_OPF} as such are rarely solved in practice. Indeed, quite often one has to tackle more advanced variants thereof. The purpose of this section is to provide a tutorial introduction to selected advanced problems that are of relevance in different operational contexts (unit commitment, ...). Moreover, we outline how the systems-and-control approaches could contribute to tackling them.

\subsection{Multi-Stage and Predictive OPF} \label{sec:MultiStage}

\begin{figure}[t]
	\centering
	\includegraphics[width=0.75\linewidth]{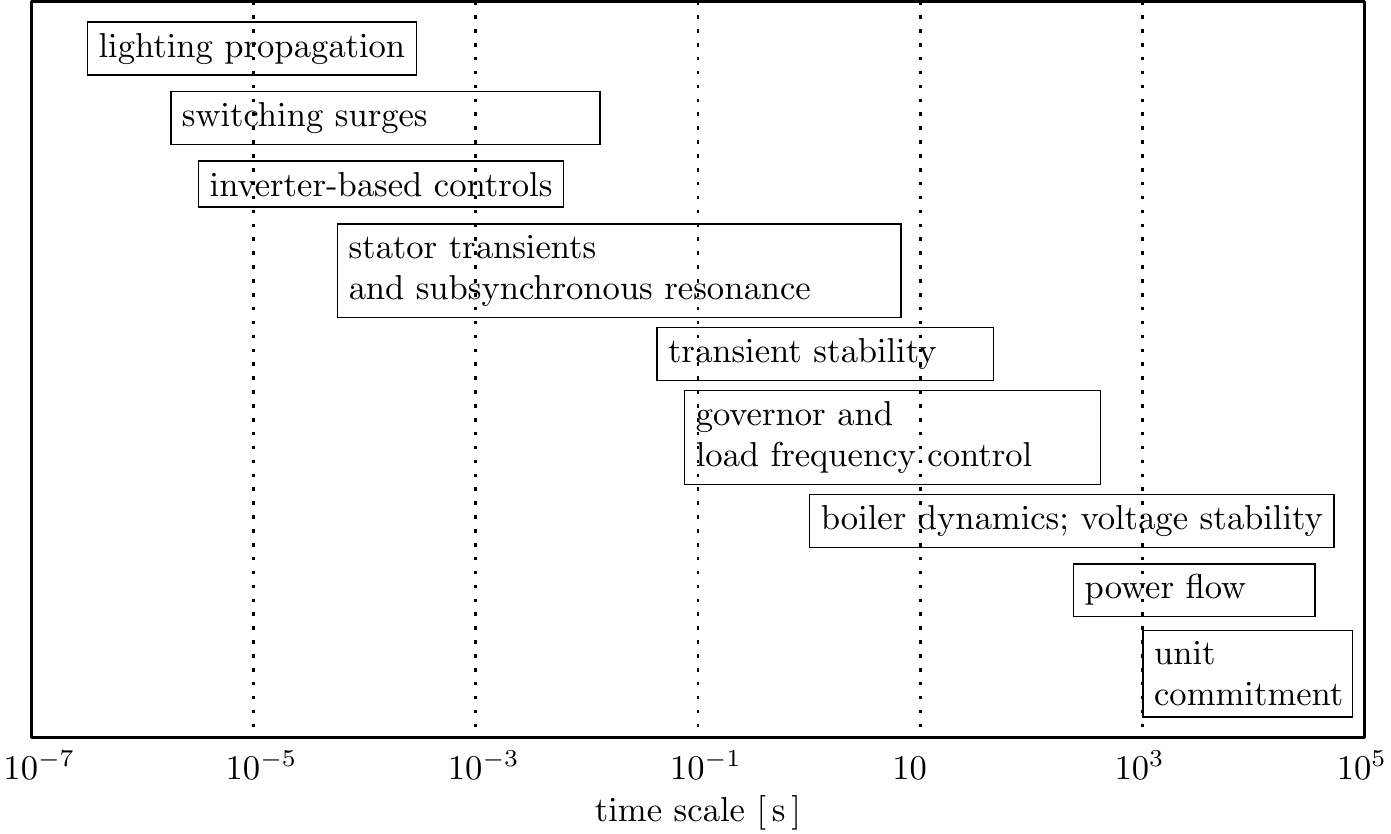}
	\caption{Time scales in power systems according to \cite{Abed2006}.}
	\label{fig:timescales}
\end{figure}

From a systems and control perspective, Problem  \eqref{eq:AC_OPF} and Problem \eqref{eq:DC_OPF} are steady-state optimization problems. Due to the well-known time scale separation underlying the conventional control paradigm of primary, secondary and tertiary control (cf. Fig. \ref{fig:timescales}), \opf problems are often solved on the basis of \unit[15]{min} sampling intervals. While the fast transients of power systems are clearly settling much faster (in the order of milliseconds up to a few minutes), a large-scale power plant cannot change its setpoint arbitrarily on a  \unit[15]{min} interval. Rather it is subject to ramp constraints that give raise to (quasi-stationary) multi-stage \opf problems. 

To the end of concise notation, henceforth the argument $\cdot(k)$ denotes the time index $k \in \mbb{N}$ of a variable. We consider ramp constraints for the generators, i.e. constraints of the form
\[
p_l(k+1) - p_l(k) \in [\Delta \underline{p}_{l}, \Delta \overline{p}_{l}], 
l \in\mcl{G}.
\]
Using the shorthand notation \eqref{eq:NodInput}, it is straightforward to see that these constraints can be expressed in 
form of the following discrete-time system
\begin{equation*}
\nodInput(k+1) = \nodInput(k) + \delta\nodInput(k), \qquad \nodInput(0) = \nodInput_0.
\end{equation*}
Here $\delta\nodInput(k) \in \mbb{R}^{n_u}$ is the incremental change of active generator powers constrained by
\[
\delta\mcl{U} := \underset{l\in \mcl{G}}{\bigtimes}\,\,  \left([\Delta \underline{p}_{l}, \Delta \overline{p}_{l}] \times
\mathbb{R}\right)\,
 \subset \mbb{R}^{n_u}.
\]
Note that the ramp constraints are typically only imposed on active power and not on reactive power.

\subsubsection{Multi-Stage AC OPF}
 The multi-stage AC \opf problem can now be stated as 
 \begin{subequations}  \label{eq:dyn_AC_OPF}
\begin{align} 
 \min_{(\nodState(\cdot), \nodInput(\cdot), \delta\nodInput(\cdot))\in\mbb{R}^{(n_x+2n_u)T}} \quad &\sum_{k \in \mcl{T}}
 J(\nodInput(k))
  + \|\delta\nodInput(k)\|_\Sigma^2 \\
 \text{subject to }& \forall k \in \mcl{T} \notag\\
 \nodInput(k+1) &= \nodInput(k) + \delta\nodInput(k), \qquad \nodInput(0) = \nodInput_0,
\\
 \delta\nodInput(k) &\in\delta\mcl{U}, \\
(\nodState(k),\, \nodInput(k))^\top &\in (\mcl{X}\times\mcl{U}) \cap \mcl{F}(\nodDist(k)), \\
\nodState(k) & \in \mcl{C},
 \end{align}
 \end{subequations}
 where $\mcl{T} = \{0, 1, \dots, k , \dots, T-1\} \subset \mathbb{N}$ denotes  the set of considered time instants.

Observe that, for the sake of generality, we introduce the quadratic penalty $ \|\delta\nodInput(k)\|_\Sigma^2$ with $\Sigma \succeq 0$ to regularize the optimization with respect to the incremental input change $\delta\nodInput(k)$. Moreover, Problem \eqref{eq:dyn_AC_OPF} looks structurally similar to discrete-time optimal control problems arising in the context of Nonlinear Model Predictive Control (\nmpc) \cite{Rawlings09, Gruene17a}. From an \nmpc point of view, $\delta\nodInput(k)$ takes the role of the input, $\nodInput(k)$ is the dynamic state, $\nodState(k)$ can be regarded as some kind of algebraic state variable, and $\nodDist(k)$ is an exogenous disturbance signal. 
Note that inclusion of energy storages (batteries, pumped-hydro, etc.) will lead to additional dynamics. Due to space limitations, we do not discuss this in detail here. Moreover, it is easy to see that Problem \eqref{eq:dyn_AC_OPF} is non-convex. However, considering the DC formulation from Section \ref{sec:AC-OPF} convex approximation is straightforward, see e.g. \cite{Hans2014, Hans2018}. 

\subsubsection*{Example -- IEEE 5 Bus}
\begin{figure}[t]
	\centering
	\includegraphics[width=0.5\linewidth]{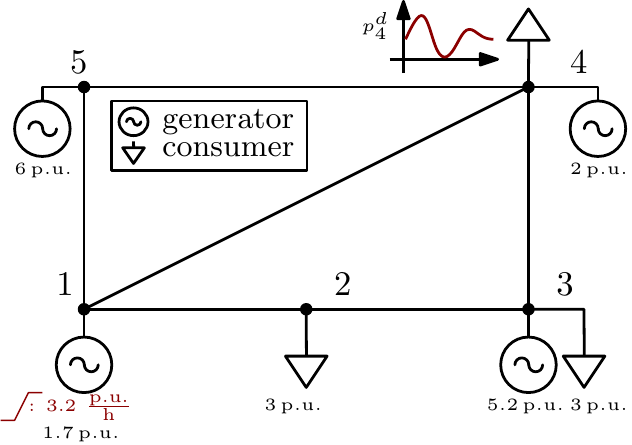}
	\caption{\ieee 5 bus system with time-varying load at node 4 and generator ramp constraint at node 1.}
	\label{fig:gridMstage}
\end{figure}

Next, we consider a simple example of a multi-stage \opf problem. Fig. \ref{fig:gridMstage} depicts the modified \ieee 5 bus case \cite{Li2010} with time-varying load at node 4 and a generator ramp constraint at the cheapest generator at node 1.
The objective $J$ is given by a quadratic function
$J(\nodInput) = \nodInput^\top H \nodInput + h^\top \nodInput, \quad H \succeq 0$,
with
$
H = \mathrm{diag}(200, 0, 220, 240, 260, \boldsymbol{0})$,  
$h = (1500, 0, 3000, 4000, 1000,\boldsymbol{0})^\top$,
where $\boldsymbol{0}$ is a vector (of appropriate dimensions) corresponding to zero cost on reactive power injection.
 Note that we do not consider regularization with respect to $\delta\nodInput(k)$.

\begin{figure}[h]
	\centering
	\includegraphics[trim={8mm 3mm 10mm 4mm},clip,width=0.65\linewidth]{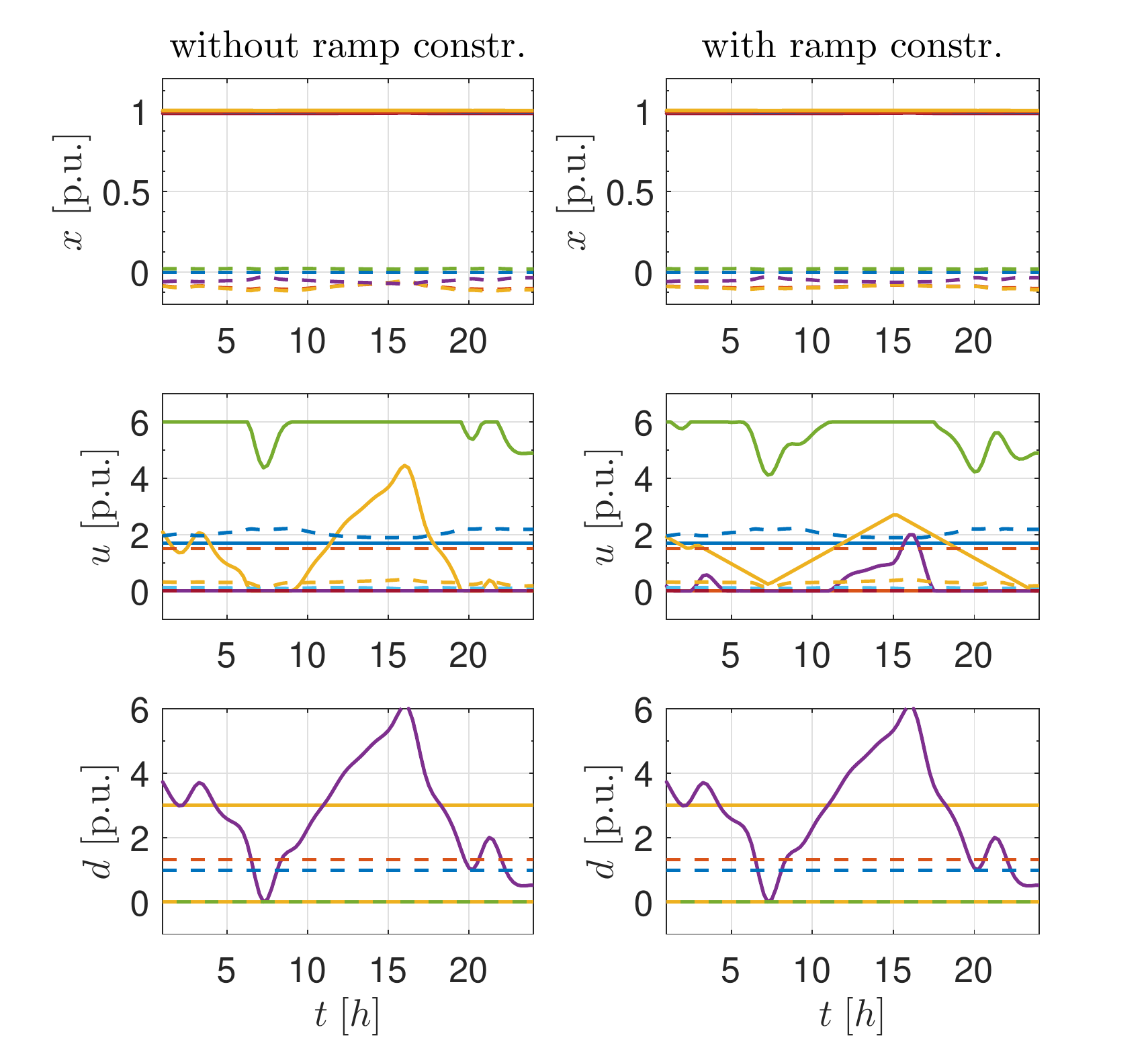}
	\caption{Results for the \ieee 5 bus case without (left) and with (right) ramp constraints.}
	\label{fig:multistageResults}
\end{figure}

Fig. \ref{fig:multistageResults} shows numerical results in case with and without active ramp constraint at node 1. We formulate the problem as an \nlp and solve it via \textsc{CasADi} and \textsc{ipopt} in \textsc{Matlab} \cite{Andersson12a}. 
Therein, the first row depicts the states consisting of phase angles $\theta$ (dashed) and voltage magnitudes $v$ (solid), the second row depicts the controls consisting of active and reactive power injections $p$ (solid) and $q$ (dashed), and the last row shows the (time varying) parameter vector $p$ consisting of active and reactive power demands $p^d$ (solid) and $q^d$ (dashed). One can observe that the time-varying demand of node 4 is mainly covered by the generators at node 1 and 5 since they are the cheapest and, hence, generator 5 is almost always on its upper limit. 
Introducing a ramp constraint of \unit[3.2]{p.u./h} for generator 1, the limited ramping capability has to be compensated by generator 3 (purple line) which has not been used in the former case. In turn this leads to higher total cost of \unit[42 930]{US\$} compared to \unit[42 166]{US\$} for the case without ramp constraints.

\subsection{Distributed OPF} \label{sec:Distributed}
Given the size of \opf problems (up to several thousands of nodes) and considering their  importance it is not surprising that there exists a large body of literature on distributed approaches to \opf, see e.g. \cite{Erseghe2015,Molzahn2017,Hug2015,Nogales2003,DallAnese2013a}.\footnote{We remark that in context  of numerical optimization for \opf problems the notions of \emph{distributed algorithms} are not unified: While in the optimization literature distributed algorithms may entail a central coordinating entity \cite{Bertsekas1989}, in context of \opf such schemes are typically referred to as being \emph{hierarchical} \cite{Molzahn2017}.} 
However, computational feasibility is not the sole motivation for distributed solutions to \opf problems. Indeed,  
distributed algorithms promise grid operation with a reduced need for centralized coordination. In other words, there is hope that  in case of blackouts or other emergencies distributed entities will show more resilience than centralized approaches \cite{Kim1997,DallAnese2013a}. 

Moreover, the current structure of the power system is inherently interconnected. For example, Germany's high-voltage transmission grid is operated by four different \tso{}s, cf. Figure \ref{fig:TSO}. On a lower level, about 890 Distribution System Operators (\dso{}s) operate the underlying distribution grids. Thus, the steadily increasing need for coordination of the different players calls for tailored numerical methods. At the same time it may be desirable to avoid accumulation of data at a single entity.

In distributed optimization one typically discusses problems with separable objectives and partially separable constraints, whereby the sole coupling is given by affine equalities \cite{Boyd2011,Bertsekas1989}; i.e. problems of the following form: 
\begin{subequations} 
\label{eq:separableForm}
\begin{align} 
\min_{(\nodState,\,\nodInput)\in\mbb{R}^{n_u+n_x}} \quad &\sum_{i\in \mathcal{R}}J_i(\nodInput_i)\\
\text{subject to}&  \notag \\
h_i(\nodState_i,\nodInput_i) &\leq 0, \qquad \quad i \in \mathcal{R}, \\
\sum_{i\in \mathcal{R}} A_i \cdot (\nodState_i&,\nodInput_i)^\top = b. \label{eq:consConstr}
\end{align}
\end{subequations}
The underlying idea is that state and control vectors $\nodState$ and $\nodInput$ are partitioned into  $\mathcal{R}=\{1,\dots,R\}$ local state and control vectors $\nodState_i\in \mathbb{R}^{n_{\nodState{}i}}$ and $\nodInput_i \in \mathbb{R}^{n_{\nodInput{}i}}$  that correspond to the subproblems.\footnote{For the sake readability, we suppress the dependence of $h_i$ on the parameter vectors $d_i$ in this section.} These subproblems involve possibly non-convex constraints $h_i: \mathbb{R}^{n_{\nodState{}i}}\times\mathbb{R}^{n_{\nodInput{}i}}\rightarrow \mathbb{R}^{n_{hi}}$.
The objective function should be a sum of local terms  depending on distinct partitions of the controls only; constraint coupling should take place via an affine so-called consensus constraint \eqref{eq:consConstr}.
Apparently, one needs to slightly reformulate Problem \eqref{eq:AC_OPF} to be consistent with Problem \eqref{eq:separableForm}. 

\subsubsection{Separable Reformulation of AC OPF}
In order to fit into the form of \eqref{eq:separableForm}, one can reformulate Problem \eqref{eq:AC_OPF} by partitioning the set of buses $\mathcal{N}$ into  $\mathcal{R}=\{1,\dots,R\}$ disjoint subsets $\mathcal{N}_i=\{n_i^1,\dots,n_i^{N_i}\} \subset \mathcal{N}$ such that $\bigcup_{i \in \mathcal{R}} \mathcal{N}_i= \mathcal{N}$ and $\mathcal{N}_i \cap \mathcal{N}_j = \emptyset$ for all $i \neq j$.
Then, one introduces two additional so-called auxiliary nodes in the middle of each line that connects two neighboring partitions. 
Thus, the node set $\mathcal{N}$ is enlarged to  $\mathcal{N}^A=\{1,\dots,N,\dots, N+A\}$ which also contains these auxiliary nodes.  
Consider an auxiliary node pair with indexes $k$ and $l$. 
In order to maintain equivalence to the original \opf problem, all values at these auxiliary node pairs should match
\begin{align*}
\theta_k &= \theta_l,&  p_k &= -p_l,\\
v_k &= v_l,&  q_k &= -q_l.
\end{align*}
Rewriting these equality constraints in matrix form yields $A_i$ and $b$ for the consensus constraint in \eqref{eq:consConstr}. Quite often, the objective of Problem \eqref{eq:AC_OPF} is the sum of squared generator powers. Hence, the objective function of \eqref{eq:separableForm} can be obtained by selecting (and possibly rearranging) the corresponding blocks to the input partitions $u_i$, cf. \cite{kit:engelmann18b} for a tutorial example.\footnote{We remark that coupling via phases, voltages \emph{and} power is just one possible problem formulation. For example in \cite{Guo2017, Erseghe2015} the coupling is enforced via voltage consensus for auxiliary nodes and their next neighbors in the interior of each region.}

Partitioning of $\mathcal{N}^A$ enables splitting the (nonlinear and non--convex) constraints  \eqref{eq:PFEconstrSet}--\eqref{eq:lineConstrSet} into local inequality  constraints $h_i$. 
To this end, the power flow equations \eqref{eq:powerflow} are formulated for all partitions $\mathcal{N}_i$ individually obtaining $\mathcal{F}_i(d_i)$ for all $i \in \mathcal{R}$. 
The same can be done for the line limits \eqref{eq:congestions} and state/input constraint sets $\mathcal{X}$, $\mathcal{U}$ obtaining $\mathcal{C}_i(d)$, $\mathcal{X}_i$, and $\mathcal{U}_i$ respectively.  

Henceforth, for the sake of compact notation, we collect all constraints forming the constraint sets on local state constraints $\mathcal{X}_i$, input constraints $\mathcal{U}_i$ and the nonlinear power flow/line flow constraints $\mathcal{F}_i(d_i)$ in one vector-valued local inequality constraint per partition 
\[
h_i(\nodState_i,\nodInput_i)\leq 0, \qquad i \in \mathcal{R}
\] 
with $h_i: \mathbb{R}^{n_{\nodState{}i}}\times\mathbb{R}^{n_{\nodInput{}i}}\rightarrow \mathbb{R}^{n_{hi}}$ completing the \opf problem in affine-coupled separable form \eqref{eq:separableForm}. 

\subsubsection{Brief Overview of Existing Approaches}

There are several research challenges for the design of distributed \opf algorithms \cite{Kargarian2017}. First of all, the algorithms should be able to solve non-convex \ac \opf problems reliably even when initialized far from a local minimum; additionally they should converge  sufficiently fast and exchange as little information as possible and, finally they should comprise a low-complexity coordination step  (ideally based on neighbor-to-neighbor communication). Furthermore, it is desirable to allow considering problem partitions related to ``operational practice''; i.e. to be able to mirror the actual \tso{}s regulation zones.

Many classical distributed optimization algorithms tailored to convex problems are---despite a lack of convergence guarantees---often applied to non-convex \ac \opf problems directly. 
Early works employ the Auxiliary Problem Principle \cite{Kim1997,Hur2002}, the Predictor Corrector Proximal Multiplier Method  \cite{Kim2000} and more recently the very popular Alternating Direction of Multipliers Method (\admm) \cite{Kim2000,Erseghe2014a}. Especially \admm has gained significant attention; thereby performing  exhaustive simulation-based convergence analysis \cite{Erseghe2014a}, investigating parameter update rules \cite{Erseghe2015}, and finally analyzing applicability to large-scale systems \cite{Guo2017}. However, for all the aforementioned algorithms convergence can, in general, not be guaranteed due to non-convexity. Moreover, the observed convergence is often slow;  especially when tight solution tolerances are needed \cite{Boyd2011,Bertsekas1989}. 

Alternatively, a method denoted as  Optimality Condition Decomposition is proposed in \cite{Conejo2002, Conejo2006} and is extended in \cite{Nogales2003, Arnold2007,Hug-Glanzmann2009}. This method performs well in many practical cases, however, there is an ongoing discussion on whether the convergence result of \cite{Conejo2002} holds for generic \ac \opf problems \cite{Erseghe2014a}.
\vspace*{2mm}

There are two major research lines dealing with the issue of convergence guarantees: 
\begin{itemize}
\item[(i)] convexifying \ac \opf by either the \dc \opf or by an (inner or outer) convex approximation of the feasible set and then applying one of the convex optimization algorithms  mentioned before; 
\item[(ii)] designing new algorithms capable of handling non-convex \ac \opf problems directly.
\end{itemize}

With respect to (i) recall that  the \ac \opf problem can be written as a problem with quadratic equality constraints, see e.g. \cite{Frank2016}. Thus it can be cast as a  rank-constrained Semi-Definite Program (\sdp) in a higher dimensional space. Dropping the non-convex rank constraint yields a convex \sdp that can be solved  with convergence guarantees \cite{Bai2008, DallAnese2013a,Molzahn2013a,Peng2018}. One limitation of the \sdp approach is that the solution obtained from the convex relaxation might not satisfy the rank constraint and hence, the solution of the original problem can not be recovered. However, there exist technical conditions (usually structural assumptions on grid components like transformers or on the grid topology) under which the exactness can be guaranteed \cite{Lavaei2012,Low2014,Low2014a}.

Research line (ii) investigates recently developed distributed optimization algorithms capable of solving non-convex problems directly with convergence guarantees. One can distinguish two main approaches: Either one distributes distinct steps of centralized algorithms like interior point or sequential quadratic programming inheriting all their mathematical properties, or one develops entirely new algorithms which directly exploit the structure of separable optimization problems. For the former, there exists approaches for distributing steps of centralized interior point methods \cite{Lu2018}, or, in the context of optimal control, there exist methods distributing steps of sequential quadratic programming \cite{Necoara2009a,TranDinh2013}. For the latter, \cite{Hours2017} presents an approach based on alternating projections combined with a trust-region globalization. Recently, a subset of the authors of the present paper proposed to use the Augmented Lagrangian Alternating Direction Inexact Newton Method (\aladin)  providing convergence guarantees and fast convergence behavior \cite{kit:engelmann17b,kit:engelmann18b}. Table \ref{tab:comparisonAll} provides a summary overview of existing distributed approaches to \opf.

\begin{table}[t]
	\caption{Overview of distributed optimization methods applied to \ac \opf.}
	\centering
	\renewcommand{\arraystretch}{1.7}
	\begin{tabular}{p{2.5cm}llp{2.5cm}l}
		\toprule
							& ADMM & ADMM-SDP & Alternating Trust Region& ALADIN  \\
		\midrule 
		Convergence \newline guarantee	& no & (yes)  & yes & yes \\ 
		Observed \newline convergence rate  	& (linear) & linear & linear & quadratic \\
		Communication \newline effort 			& low &  low/medium   & low/medium & medium/high \\
		Selected \newline references  	& \cite{Erseghe2014a,Guo2017} & \cite{DallAnese2013a} & \cite{Hours2017} & \cite{kit:engelmann17b,kit:engelmann18b} \\	
		\bottomrule		
	\end{tabular}
	\label{tab:comparisonAll}
\end{table}

\subsubsection{Example -- IEEE 14 Bus via ALADIN} \label{sec:Example}

As an example for solving \ac \opf \eqref{eq:AC_OPF} in a hierarchical distributed fashion, we apply the \aladin algorithm to the \ieee 14 bus test system. We start by briefly recalling \aladin, for a more detailed description of \aladin we refer to \cite{Houska2016}.

In step 1) of Algorithm \ref{ALADINAlg}, the local \nlp{}s \eqref{eq:localNLP} are solved obtaining local optimal inputs and state vectors $z_i = (\nodState_i,\nodInput_i )$
for all $i \in \mathcal{R}$. In these \nlp{}s an augmented Lagrangian function is minimized with respect to the consensus constraint of Problem \eqref{eq:separableForm}. The Lagrange multipliers $\lambda^k$ and $\bar z_i^k$ are treated as fixed parameters.
Step 2) computes local derivatives; i.e. gradients $g_i^k$, Hessians $H_i^k$ and Jacobians $C_i^k$ of the active constraints.
The index set of active constraints (inequality constraint holding with equality) at $(z_i^k)$ is defined by
$
\mathcal{A}_i(z_i^k) = \{\; j\;\, |\;\, (h_i)_j(z_i^k) = 0\;\}
$
and communicated to the coordinating entity. Observe that in many cases these derivatives do not have to be computed explicitly as they are returned by the solvers applied to  \eqref{eq:localNLP}. In the following step 3), all derivatives from the local 
problems are collected, aggregated in global derivatives $g^k$, $H^k$ and $C^k$, and an equality constrained coordination \qp is solved. This is computationally cheap as solving this \qp leads to a linear system of equations where very efficient solvers exist. Finally, step 4) updates the solution guesses and performs a line search if necessary (this step can be omitted in many cases, cf. \cite{kit:engelmann18b}). The very last step applies a problem-specific heuristic to update $\rho$ and $\mu$.

\begin{algorithm2e}
	\caption{\aladin algorithm.  } \label{ALADINAlg}
	\SetAlgoLined
	\KwResult{$z^\star$ }
	\textbf{Input:}$\quad \bar z^0, \; \lambda^0, \; \rho, \; \Sigma_i, \; \mu, \;$ $k=0$ \vspace*{.25cm}\\
	\While{$\left \|A   z^k \right \|_\infty > \epsilon$ \textbf{and} $\rho \, \left \|  z^k  -\bar z^k \right \|_\infty > \epsilon$ \vspace*{.1cm}}{

		\textbf{\label{step1}1) Solve local problems}
		\begin{align}\notag
		z_i^{k} = \underset{z_i}{\text{arg}\min}&\quad f_i(z_i) + (\lambda^k)^\top A_i  z_i + \frac{\rho^k}{2}\left\|z_i -\bar z_i^k \right\|_{\Sigma_i}^2\quad\\
		\text{s.t.}\quad&\quad  h_i(z_i) \leq 0 \quad \mid \kappa_i^k.		\label{eq:localNLP}
		\end{align}

		\textbf{2) Compute gradients \& Hessians for QP}

		Obtain gradients $g_i^k$, Hessian of the Lagrangians $B_i^k$ the Jacobian of the active constraints $C_i^k$

		\textbf{3) Solve coordination QP}
		Solve the coordination \qp
		\begin{align}
		\notag
		&\underset{\Delta z_i, s}{\min}\;\;\sum_{i\in \mathcal{R}}\left\{\frac{1}{2}\Delta z_i^\top B^k_i\Delta z_i + {g_i^k}^\top \Delta z_i\right\}     + (\lambda^k)^\top s + \frac{\mu^k}{2}\|s\|^2_2  \\
		&
		\begin{aligned}\label{eq::conqp}
		\qquad \qquad \text{s.t.}\; \qquad  \qquad \quad C^k \Delta z_i  &= 0 \\
		\quad \sum_{i\in \mathcal{R}}A_i( z_i^k + \Delta z_i) &=  s     \quad |\; \lambda^\mathrm{QP},
		\end{aligned}
		\end{align}
		obtaining $\Delta z^k$ and $\lambda^{\text{QP}}$.

		\textbf{4) Line search}
		Update primal and dual variables by
		\begin{eqnarray}\notag
		\bar z_i^{k+1} &\leftarrow& z_i^{k} + \alpha^k_1 \left ( z_i^{k} -\bar z_i^{k} \right ) + \alpha_2^k  \Delta z_i^{k} \;,\\[0.2cm]\notag
		\lambda^{k+1}&\leftarrow&\lambda^k + \alpha^k_3 (\lambda^\mathrm{QP}-\lambda^k),
		\end{eqnarray}
		with $\alpha^k_1,\alpha^k_2,\alpha^k_3$ from~\cite{Houska2016}. If full step
		is accepted, i.e. $
		\alpha_1^k=\alpha_2^k=\alpha_3^k=1,
		$
		update $\rho^k$ and $\mu^k$ by
		\begin{align}
		\rho^{k+1}\;(\mu^{k+1}) =
		\begin{cases}
		r_\rho \rho^k\;(r_\mu \mu^k)   &\text{if} \; \rho^k < \bar \rho\;\; (\mu^k < \bar \mu)\\ \nonumber
		\rho^k\;(\mu^k)  &\text{otherwise}
		\end{cases} .
		\end{align}
	}
\end{algorithm2e}

Conceptually, \aladin  
can be regarded as a combination of \admm and an \sqp method; i.e. the local steps are very similar to the ones of \admm, but the coordination step is similar to \sqp. Furthermore, 
it is possible to express \admm as a special case of \aladin by appropriate choice of parameters \cite{Houska2016}.

\begin{figure}[t]
	\centering
	\includegraphics[width=0.45\textwidth]{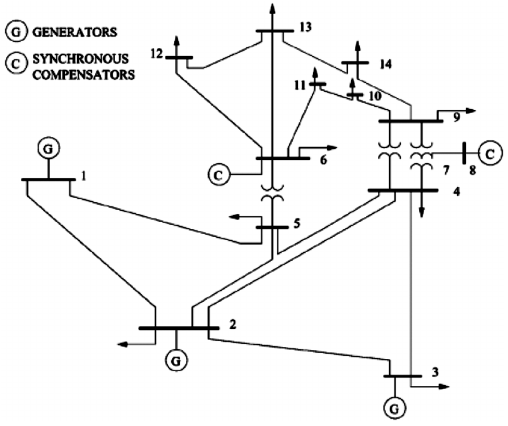}
	\caption{\noindent IEEE 14 bus test system with $\mathcal{N}_1=\{1,2,5\}$, $\mathcal{N}_2=\{3,4,7\text{\,-\,}9\}$, $\mathcal{N}_3=\{6,10\text{\,-\,}14\}$. }
	\label{fig:14-busMap}
\end{figure}

\begin{figure}[h]
	\centering
	\includegraphics[trim={20mm 0 23mm 0},clip,width=1\textwidth]{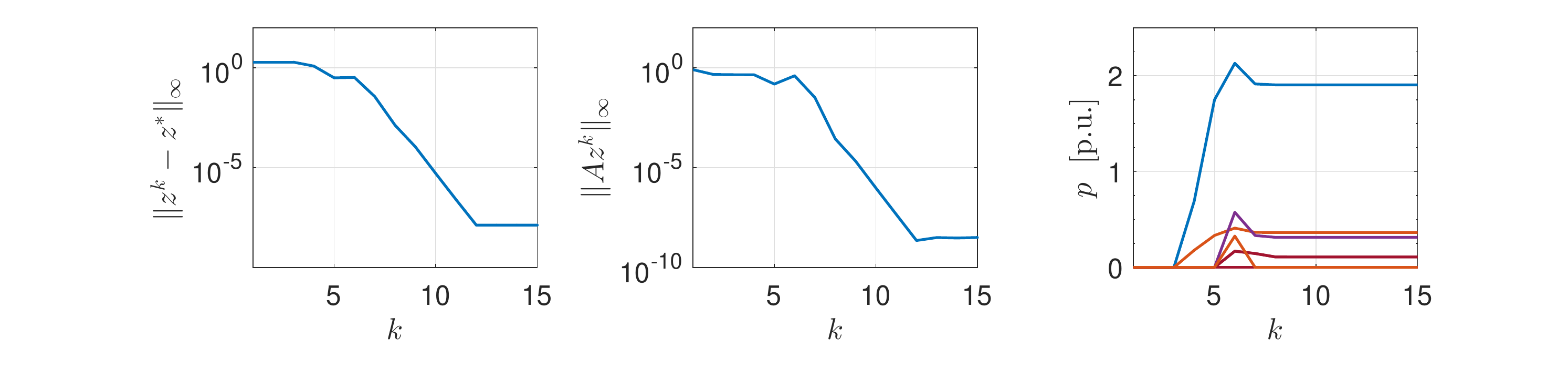}
	\caption{\noindent Numerical results for the \ieee 14 bus test system.}
	\label{fig:14-busResults}
\end{figure}

We consider the \ieee 14 bus test system shown in Fig. \ref{fig:14-busMap} as an illustrative example. We divide the grid into $\mathcal{R}=\{1,2,3\}$ partitions motivated by geographical considerations. Due to space limitations,  we omit commenting on the  parameter tuning for \aladin here. We refer to \cite{kit:engelmann18b,kit:engelmann17b} for an \opf--specific discussion.
We solve the problem via \textsc{CasADi} and \textsc{ipopt} in \textsc{Matlab} \cite{Andersson12a}.
  Fig. \ref{fig:14-busResults} shows the numerical results of \aladin for the described test system. Specifically, we plot the distance to the ``true'' minimizer $\| z^k-z^\star\|_\infty$, the consensus violation $\|A  z^k\|$---which indicates to which extent the 
physical values (active/reactive power and voltages) at the auxiliary nodes match---and the active power injections $p$ (which partially represents the controls $u$) over the iteration index $k$. For all these indicators, \aladin converges to high accuracy in less than 15 iterations. Compared to \admm, this is significantly faster: \admm usually needs at least around hundred iterations to attain  medium accuracy for problems of similar size, cf. \cite{Erseghe2015,kit:engelmann18b}.

However, note that the complexity per iteration of \aladin compared to \admm is higher. Whereas \admm exchanges local solution guesses $z_i^k$ only, \aladin additionally communicates derivatives of the objective and the constraints which increases the per step communication need. Furthermore, the coordination step is more complicated because \aladin requires the solution of a linear system of equations whereas for \admm this system can be reduced to the computation of averages \cite{Boyd2011}.

\subsection{OPF with Uncertainties} \label{sec:Stochastic}

Traditionally, the \opf problems~\eqref{eq:AC_OPF} and \eqref{eq:DC_OPF} are solved for a fixed value of the disturbance $\nodDist \in \mathbb{R}^{n_{\nodDist}}$.
However, demand forecasting and the feed-in of renewable energy sources---to name just a few drivers---call for a structured consideration of uncertainties.
In the presence of uncertainties it may be more adequate to model uncertain feed-in and/or uncertain demand by random variables.\footnote{Robust \emph{min-max} solutions are typically not favored in practice.  
They imply consideration of worst-case scenarios and may lead to high operational costs. Moreover, the underlying assumptions are hard to verify and thus theoretical guarantees may not hold in reality. }
Importantly, \opf-specific uncertainties can be modeled by Gaussian \emph{and} non-Gaussian random variables \cite{Atwa10a,Carpaneto08a,Soubdhan09a}.

\subsubsection{Conceptual Considerations}
To the end of formalizing the \opf with random variables, we consider the Hilbert space of random variables of finite variance.
For a given set of outcomes $\Outcomes$, a $\sigma$-algebra $\mathcal{A}$, and a probability measure $\ProbMeasure$, let $\ProbSpace$ denote the corresponding probability space.
The space
\begin{equation}
\label{eq:L2space}
\Ltwospace{} = \left\{ \rv{x}: \Outcomes \rightarrow \mathbb{R} \quad | \quad \int_{\Outcomes} \rv{x}(\tau)^2 \mathrm{d} \ProbMeasure (\tau)  \right\} ~ / ~ \ProbMeasure\text{-almost everywhere}
\end{equation}
is the set of all equivalence classes of random variables of finite variance \cite{Sullivan2015}.
The space $\Ltwospace{}$ according to~\eqref{eq:L2space} is a Hilbert space with respect to the scalar product
\begin{equation*}
\langle \rv{x}, \rv{y} \rangle = \ev{\rv{x} \rv{y}} = \int_{\Outcomes} \rv{x}(\tau) \rv{y}(\tau) \mathrm{d} \ProbMeasure (\tau).
\end{equation*}
By slight abuse of terminology we refer to the equivalence classes simply as ``random variables.''
For $\mathbb{R}^{n_{\nodState}}$-valued random vectors the following notation is introduced
\begin{equation*}
\rv{x} \in \Ltwospace{n_{\nodState}} \quad \Longleftrightarrow \quad \rv{x}_i \in \Ltwospace{} \quad \forall i = 1, \hdots, n_{\nodState}.
\end{equation*}
That is, instead of treating the disturbance $\nodDist$ as a real-valued element of $\mathbb{R}^{n_{\nodDist}}$, we view it as a random vector $\rv{\nodDist} \in \Ltwospace{n_{\nodDist}}$.
Note that so far no specific kind of distribution is assumed for the disturbance; the probability measure $\ProbMeasure$ could be Gaussian, but could also refer to any other kind of distribution of finite variance.
Importantly, multivariate distributions can be considered in case the Hilbert space $\Ltwospace{}$ is viewed as a tensor product space of appropriate univariate Hilbert spaces \cite{Sullivan2015}.
We continue by investigating the consequences for the \opf problem~\eqref{eq:AC_OPF} in case of modeling the disturbance as a random variable.

First, the power flow equations~\eqref{eq:PFE_compact} will almost surely be (numerically) violated for any combination of states $\nodState\in \mathbb{R}^{n_{\nodState}}$ and control inputs $\nodInput\in \mathbb{R}^{n_{\nodInput}}$, because
\begin{equation*}
F: \mbb{R}^{n_{\nodState}} \times \mbb{R}^{n_{\nodInput}} \times \Ltwospace{n_{\nodDist}} \rightarrow \mbb{R}^{2 N}: \quad
F(\nodState, \nodInput; \rv{\nodDist}) \neq 0 \quad \text{a.s.} ~.
\end{equation*}
From the numerical point of view, in the presence of uncertainties the power flow equations will almost surely be violated if the state $\nodState$ and the input $\nodInput$ are fixed real-valued quantities.
However, even if the numerical solution $ (\nodState,\, \nodInput)\in \mathbb{R}^{n_{\nodState} + n_{\nodInput}}$ violates the power flow equations, the real physical system---obeying the laws of physics---of course attains a state on the power-flow manifold corresponding to the specific input $\nodInput\in \mathbb{R}^{n_{\nodInput}}$.
Any numerical deviations of the power flow equations have to be accounted for by lower-level controllers.
In the view of hierarchical control of power systems it is  thus desirable to obtain higher-level control inputs $\nodInput$ from \opf such that the numerical solution is as close to the physical solution as possible; this ensures fewer control actions (of lower magnitude) to be taken at the lower levels.

A possible way of achieving this is to consider the power flow equations as a nonlinear mapping from random variables to random variables, i.e. $F: \Ltwospace{n_{\nodState}} \times \Ltwospace{n_{\nodInput}} \times \Ltwospace{n_{\nodDist}} \rightarrow \Ltwospace{2 N}$ with
\begin{equation}
\label{eq:PFE_RV}
F(\rv{\nodState}, \rv{\nodInput}; \rv{\nodDist}) = 0 \quad \Longleftrightarrow \quad \forall \omega \in \Outcomes: ~ F(\rv{\nodState}(\omega), \rv{\nodInput}(\omega); \rv{\nodDist}(\omega)) = 0,
\end{equation}
and consequently
\begin{equation*}
\mcl{F}(\rv{\nodDist}) := \left\{(\rv{\nodState} ~ \rv{\nodInput})^\top \in \Ltwospace{n_{\nodState}} \times \Ltwospace{n_{\nodInput}} \,|\, F( \rv{\nodState}, \rv{\nodInput}; \rv{\nodDist}) = 0  \right\} .
\end{equation*}
In other words,  every outcome $\omega \in \Outcomes$ corresponds to a triple $(\rv{\nodState}(\omega), \rv{\nodInput}(\omega), \rv{\nodDist}(\omega))$ comprised of a realization of the input $\rv{\nodInput}(\omega)$, a realization of the state $\rv{\nodState}(\omega)$, and a realization of the disturbance $\rv{\nodDist}(\omega)$. Importantly, this triple should satisfy  the power flow equations; in \cite{Bienstock2014} this notion is referred to as \emph{viability}.
Viability of the power flow equations can thus be ensured by formally introducing random variables $\rv{\nodState}$ and $\rv{\nodInput}$ for the state and the input, respectively.
This way, similar to well-known \textsc{lqg} control, any viable random-variable input $\rv{\nodInput}$ corresponds to a feedback policy with known probabilities of certain control actions to be taken.

\subsubsection{Stochastic OPF}
In the presence of stochastic uncertainty surrounding the disturbance $\nodDist$, the goal of \opf is to compute optimal viable feedback policies $\rv{\nodInput}$---this problem is called stochastic \opf (\sopf).
Viability can be ensured by enforcing the random-variable power flow equations~\eqref{eq:PFE_RV} as equality constraints.
As \sopf optimizes over policies~$\rv{\nodInput}$, its cost function has to map policies to scalars
\begin{equation*}
\hat{J}: \Ltwospace{n_{\nodInput}} \rightarrow \mathbb{R},
\end{equation*}
e.g. $\hat{J}(\rv{\nodInput}) = \ev{J(\rv{\nodInput})}$.
Additionally, the inequality constraints from \opf \eqref{eq:AC_OPF} have to be adjusted, because inequalities in terms of random variables are not meaningful in general.
This can be achieved, for example, by introducing chance constraints.
A possible formulation for \sopf using joint chance constraints for the inequality constraints reads
\begin{subequations}
	\label{eq:AC_OPF_stochastic}
	\begin{align} 
	\min_{(\rv{\nodState}, \rv{\nodInput})   \in\Ltwospace{n_{\nodState} + n_{\nodInput}}} \quad & \hat{J}(\rv{\nodInput})  \\
	\text{subject to}\qquad \qquad &  \notag\\
	(\rv{\nodState} ~ \rv{\nodInput})^\top &\in \mcl{F}(\rv{\nodDist}), \\
	\ProbMeasure \left( \rv{\nodInput} \in \mcl{U} \right) &\geq 1 - \varepsilon_{\nodInput}, \\
	 \ProbMeasure \left( \rv{\nodState} \in \mcl{X} \right) &\geq 1 - \varepsilon_{\nodState}, \\
	 \ProbMeasure \left( \rv{\nodState} \in \mcl{C} \right) &\geq 1 - \varepsilon_{c},	
	\end{align}
\end{subequations}
where $\varepsilon_{\nodInput}, \varepsilon_{\nodState}, \varepsilon_c \in [0, 1]$ are risk levels specified by the user.
It is worth to be noted that any feasible solution to Problem~\eqref{eq:AC_OPF_stochastic} 
is a viable feedback policy satisfying the inequalities in the prescribed chance-constrained sense.

\subsubsection{Brief Overview of Existing Approaches}
There exist various reformulations of \opf in the presence of uncertainties: for example \cite{Roald15,Roald15b,Bienstock2014} employ individual chance constraints, 
\cite{Vrakopoulou12,Warrington13} use joint chance constraint reformulations and extend the setting to the multistage setting, and \cite{Muehlpfordt16b,kit:Muehlpfordt17b} formulate the problem entirely in terms of random variables.
The reformulation of the cost function and the inequality constraints in the presence of uncertainties is neither unique nor is there consensus in the literature on which one is more preferable than another.

It remains to address how to solve the infinite-dimensional \sopf problem \eqref{eq:AC_OPF_stochastic}.
For example, it is possible to solve the chance-constrained optimization problem by means of multi-dimensional integration \cite{Zhang11,Zhang13}.
However, by this approach no explicit feedback policies are obtained.
Other lines of research hence focus on reformulating the chance constraints and parameterizing the infinite-dimensional decision variable to obtain easier-to-solve deterministic finite-dimensional optimization problems.
For reformulations of chance constraints in the context of power systems see \cite{Roald15b}; we refer to \cite{Popescu05,Calafiore13a} for more general references.
Affine parameterizations of the feedback policies are popular for \sopf, especially in the context of \dc power flow \cite{Vrakopoulou13,Roald13,Roald15,Roald15b,Bienstock2014}.
How to convert the infinite-dimensional \sopf problem \eqref{eq:AC_OPF_stochastic} to a finite-dimensional deterministic problem is shown, for example, in \cite{Bienstock2014,Muehlpfordt2018b}; it is also shown there that for \dc power flow conditions affine policies are always viable.
We remark that so far it remains an open question as to what kind of feedback policies are viable for generic \ac \opf problems subject to uncertainties.

Due to space limitations, we do not provide an example on \sopf using the sketched $L^2$ approach. Instead we refer to the ones provided in \cite{kit:Muehlpfordt17b, Muehlpfordt2018b, kit:muehlpfordt18b}.

\subsection{Further OPF Variants}

In the preceding sections, we discussed different variants of the \opf problem. Yet the list of variants discussed is not exhaustive. Specifically, we have not been touching upon secure \opf variants and variants with discrete decision variables. 

The $N-1$ security constrained \opf problem takes into account the requirement that the system should withstand the loss of a single component (which could be a transmission line, a generator, a transformer etc.) \cite{Glanzmann2006, Cao2013,Capitanescu2011}. For the sake of simplicity, one often considers the most important case of transmission line outages whereby usually \dc-approximations of the power flow equations are used \cite{Zhu2015}.

Moreover, several controllable devices in power systems such as, e.g., transformer settings can only take discrete (input) values. Thus, discrete decision variables arise frequently in \opf. For example, optimal reactive power dispatch is a variant of \opf, where the active power injections $p$ are assumed to be given (e.g. by an energy market) and only the remaining variables which mainly are the reactive power injections, transformer settings, shunt settings or \facts set-points are used for minimizing the losses in the grid \cite{kit:murray18c}.

\section{Open Research Questions} \label{sec:openProbs}
Given the \ac \opf Problem \eqref{eq:dyn_AC_OPF}  and its variants mentioned in Section \ref{sec:Challenges}  it remains to sketch open research challenges for systems and control. We will first comment on the challenges of the individual subproblems and then we turn towards the overarching ones. 

\subsection{Multi-Stage OPF and NMPC}
Recalling that the multi-stage \opf  \eqref{eq:dyn_AC_OPF} is a discrete-time optimal control problem, it is evident that \nmpc appears to be a promising means of solving it in receding-horizon fashion. Indeed, several works have already suggested doing so \cite{Hiskens2005,Gill2014,Meyer-Huebner2015}. However, important conceptual developments of \nmpc---in term sof stability analysis and recursive feasibility analysis---have not yet been transferred to \opf. For example, it remains open how to transfer the considerable existing body of knowledge on efficient real-time iteration schemes for \nmpc, which originally has been developed having in mind process control applications, to Problem \eqref{eq:dyn_AC_OPF}? More precisely, what kind of computational performance can be expected from real-time iteration schemes such as \cite{Diehl05a, Diehl05b, Wolf16a} when applied to large-scale \opf problems? 

From a systems theory perspective, we note that the objective of  \eqref{eq:dyn_AC_OPF} is not a conventional tracking term---i.e. it is not a distance to some pre-computed setpoint. Hence, the receding-horizon solution of \eqref{eq:dyn_AC_OPF} falls into the realm of \emph{economic \nmpc} \cite{kit:faulwasser18c}. Therefore, it is fair to ask whether one needs to a add stabilizing constraints / terminal penalties to  \eqref{eq:dyn_AC_OPF} to enforce convergence and stability in an economic \nmpc framework? Or shall one avoid those constraints and rather analyze  \eqref{eq:dyn_AC_OPF} in a notion of time-varying turnpike properties \cite{kit:faulwasser18c,Gruene17b}?
From a power systems point of view, it would be interesting to analyze how to combine efficient online optimization with approaches of alternating optimization and power-flow solution and how to encode additional application requirements (e.g. maximal islanding time, black start, line switches, ...). With respect to the later issue elements of a convex reformulation of maximal-islanding-time constraints are presented in \cite{kit:braun18a}.
Finally, the fact that for $k \in \mcl{T}$ the non-convex constraint set $ \mcl{F}(\nodDist(k))$ depends on future (hence uncertain) values of $\nodDist(k)$ underpins that uncertainty is always pivotal in \opf Problems. 

\subsection{Stochastic and Distributed OPF}
With respect to \opf subject to uncertainties, Section \ref{sec:Stochastic} has highlighted the conceptual promise of working with random variables. However, as soon as one moves from single-stage to multi-stage \opf, the time-wise correlation of uncertainties poses considerable conceptual challenges. Put differently, at the present stage it is unclear how to design stochastic \nmpc for multi-stage \ac \opf such that the power-flow constraints are viably satisfied. Even in the single-stage case it is not yet clear how to transfer the \dc results on viable formulations of \cite{kit:Muehlpfordt17b,kit:muehlpfordt18b} to the \ac setting. Finally, it is also worth investigating how to solve \sopf in distributed fashion. To this end, the combination of \aladin with polynomial chaos   is investigated in \cite{kit:engelmann18a} for small scale problems. 

In  context of distributed \opf  there remain open issues with respect to the applicability of \sdp relaxations, with respect to a trade-off between communication effort and number of iterations, and with respect to the interplay of grid partitioning and convergence properties. 

\subsection{Towards Flexible Energy Cells with Partial Autonomy}
From the application point of view, advanced optimization-based control of energy systems promises a structured approach towards clustering, design, and operation of (partially) autonomous subsystems (so-called energy cells), which is regarded as one potentially viable option for the future, cf. \cite{cSells}. Put differently, one is interested in operating subsystems in a flexible manner such that they can be either coupled to an upper-level grid and/or such that they may be temporarily disconnected if needed.
 
 First results on scheduling of systems  combining \res and energy storage show promising performance while they neglect the underlying grid topology \cite{kit:appino17b,kit:appino18b}. Moreover, the need to rely on data-driven forecasts underpins once more the promise of investigating the confluence of data-driven machine-learning and control. 

From a systems-and-control point of view it is tempting to address the design of energy cells by means of investigating distributed stochastic economic \nmpc of energy systems. While first results on  distributed economic \nmpc  \cite{Kohler17a} and on scenario-based  \nmpc \cite{Hans15a} have been presented, an exhaustive analysis is still open. In this context, key aspects will be the understanding of how to construct data-driven forecasting schemes for power respectively energy and how to utilize these forecasts for scheduling and control.
However, it is evident that any advanced and/or predictive control of energy systems will have to rely on control-oriented modeling and will require applicable solutions for the dual (decentralized/distributed) state estimation problem.

\section{Conclusions} \label{sec:Conclusions}

Accelerated by the Energiewende, operation and control of electrical energy systems have to deal with the increasing in-feed of renewable energy. 
This leads to a certain vulnerability of the stability of the energy system owing to increased volatility of generation and, at the same time, leading to the reduction of inertia of the large generators in the system.
Moreover, there has been a long time span without widespread interaction between Electrical Engineering, Control Engineering and Mathematical Systems Theory leading to a lack of common vocabulary.

On this canvas, the present paper focused on Optimal Power Flow
problems that are of tremendous importance for power systems and that arise in several contexts of operation of such systems. Starting from a brief introduction to \ac and \dc \opf, we presented three challenging variants in a unified framework: multi-stage \opf, distributed \opf, and \opf with uncertainties. Furthermore, we presented case-study examples for multi-stage and distributed \opf. 

Aiming to foster the interaction between systems-and-control community and the power-systems community, we commented on open research problems that might be of interest for the systems-and-control community. We remark that the optimization-based methods sketched here necessitate concurrent progress on the underlying (fast time-scale) problems of voltage and frequency stabilization via distributed and decentralized control. 

Finally, it is clear that the Energiewende provides a plethora of highly relevant research challenges for systems and control. Indeed the transition can only be successful if the ties between communities are strengthened.

\printbibliography

\end{document}